\documentclass[amsmath, amsthm, amssymb, aps, prb, superscriptaddress, twocolumn, nofootinbib, 10pt]{revtex4-1}
\usepackage{setspace}
\usepackage{amsmath}
\usepackage{breqn}
\usepackage{graphicx}
\usepackage[nearskip,margin = 0pt,caption=false]{subfig}

\usepackage{verbatim}
\usepackage{amsfonts}
\usepackage{amssymb}
\usepackage{epstopdf}
\usepackage{lipsum}
\usepackage{siunitx}
\usepackage{xcolor}
\usepackage{array}
\usepackage{braket}
\usepackage[normalem]{ulem}
\usepackage[resetlabels,labeled]{multibib}
\DeclareGraphicsExtensions{.pdf,.eps,.png,.jpg,.mps}
\usepackage{etoolbox}
\usepackage{subfiles} 

\begin{document}
\title{Titanium:Sapphire-on-insulator for broadband tunable lasers and high-power amplifiers on chip}
\author{Joshua Yang$^{\dagger}$, Kasper Van Gasse$^{\dagger}$, Daniil M. Lukin$^{ \dagger}$,  Melissa A. Guidry, Geun Ho Ahn, Alexander D. White,  Jelena Vu\v{c}kovi\'{c}$^{*}$\\
\vspace{+0.05 in}
E. L. Ginzton Laboratory, Stanford University, Stanford, CA 94305, USA.\\
{\small $^{\dagger}$These authors contributed equally to this work.}\\
{\small $*$ jela@stanford.edu}}

\begin{abstract}
\noindent 
Titanium:Sapphire lasers have been essential for advancing fundamental research and technological applications, including the development of the optical frequency comb\cite{holzwarth2000optical}, two-photon microscopy\cite{helmchen2005deep} and experimental quantum optics\cite{ semeghini2021probing, ebadi2021quantum}.
Ti:Sapphire lasers are unmatched in bandwidth and tuning range, yet their use is severely restricted due to their large size, cost, and need for high optical pump powers\cite{moulton1986spectroscopic}. 
Here, we demonstrate a monocrystalline Titanium:Sapphire-on-insulator (Ti:SaOI) photonics platform which enables dramatic miniaturization, cost-reduction, and scalability of Ti:Sapphire technology.
First, through fabrication of low-loss whispering gallery mode resonators, we realize a Ti:Sapphire laser operating with an ultra-low lasing threshold of 290~\textmu W. 
Then, through orders-of-magnitude improvement in mode confinement in Ti:SaOI waveguides, we realize the first integrated solid-state (i.e., non-semiconductor) optical amplifier operating below 1~\textmu m, with an ultra-wide bandwidth of 700 - 950~nm and peak gain of 64~dB/cm. We demonstrate unprecedented 17~dB distortion-free amplification of picosecond pulses to up to 2.3~nJ pulse energy, corresponding to a peak power of 1.0~kW. 
Finally, we demonstrate the first tunable integrated Ti:Sapphire laser, featuring narrow linewidths and a 24.7~THz tuning range, which, for the first time, can be pumped with low-cost, miniature, off-the-shelf green laser diodes. 
This opens doors to new modalities of Ti:Sapphire lasers (now occupying a footprint less than 0.15~mm\textsuperscript{2}), such as massively-scalable Ti:Sapphire laser array systems for a variety of applications. As a proof-of-concept demonstration, we employ a Ti:SaOI laser array as the sole optical control for a cavity quantum electrodynamics experiment with artificial atoms in silicon carbide\cite{lukin2023two}. 
This work is a key step towards the democratization of Ti:Sapphire technology through a three orders-of-magnitude reduction in cost and footprint, as well as the introduction of solid-state broadband amplification of sub-micron wavelength light.

\end{abstract}

\maketitle

\section{Introduction} 

%
%
\begin{figure*}[t!]
\includegraphics[width=\linewidth]{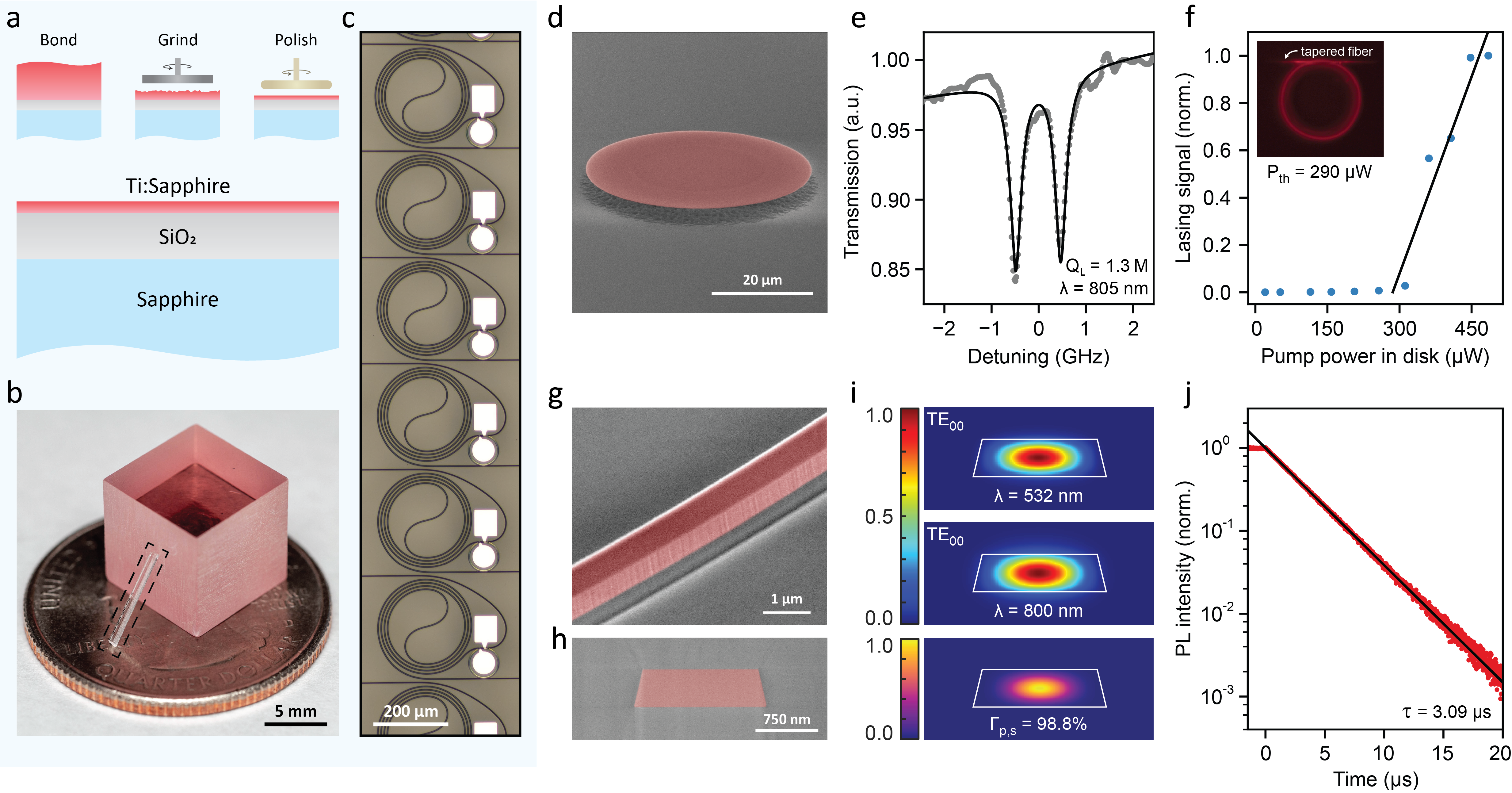}
\centering
\caption{\textbf{Low-loss photonics and sub-milliwatt threshold Ti:Sapphire laser in monocrystalline sapphire-on-insulator. }
(a) Process flow for Ti:SaOI preparation. 
(b) Photograph of an array of on-chip Ti:SaOI lasers (inside dashed rectangle) resting on a Ti:Sapphire bulk crystal, and (c) a close-up optical micrograph of fabricated devices. 
(d) Scanning Electron Microscope (SEM) image of a photolithographically-patterned Ti:SaOI microdisk resonator (Ti:Sapphire layer is false colored).
(e) Quality factor of $1.3\times10^6$ measured in transmission at 805~nm, fit to a split Lorentzian.
(f) Lasing signal of the microdisk resonator (measured via spectrometer) with increasing pump power in the resonator, revealing lasing threshold of 290~\textmu W. Inset: optical microscope image of the resonator under excitation (with 532~nm light filtered), showing spontaneous emission collected vertically.
(g) SEM image of a Ti:SaOI waveguide fabricated with electron-beam lithography. 
(h) SEM of waveguide cross-section. 
(i) Simulated mode intensity profile at the pump (532 nm) and lasing (800 nm) wavelength, with modal intensity overlap $\Gamma_{p,s}$ of 98.8\%. 
(j) Exponential decay of waveguide fluorescence  with an optical lifetime of 3.09~\textmu s.
}
\label{fig_1}
\end{figure*}

Titanium:Sapphire (Ti:Sapphire) laser systems \cite{moulton1982ti, moulton1986spectroscopic} play an essential role in both fundamental research and technological applications, including the demonstration of the first optical frequency comb \cite{holzwarth2000optical}, generation of optical pulses as short as two oscillations of the electrical field \cite{morgner1999sub}, two-photon microscopy and optogenetics \cite{helmchen2005deep, prakash2012two, xu1996measurement}, and demonstrations of on-chip laser driven particle accelerators \cite{leedle2015laser, kfir2020controlling}. 
As a solid-state (i.e., non-semiconductor) gain medium, Ti:Sapphire has exceptional properties, possessing the widest gain bandwidth of any laser crystal (650 - 1100 nm), a large emission cross-section, and a four-level structure \cite{moulton1986spectroscopic}. Consequently, mode-locked and continuous-wave Ti:Sapphire lasers are unmatched in performance, and are essential in disciplines such as quantum optics and atomic physics \cite{hayes2010entanglement, santori2002indistinguishable, rugar2021quantum, semeghini2021probing, ebadi2021quantum}. 
However, the unrivaled performance of Ti:Sapphire lasers comes at a cost: commercial systems are bulky, expensive, and require high-power pump lasers, prohibiting their widespread use in many real-world applications that demand compactness and scalability. 

Photonic integrated circuit technologies are revolutionizing laser systems by delivering compactness, scalability, and cost-efficiency that is impossible to achieve in table-top systems, leveraging wafer-scale fabrication capabilities in material platforms such as silicon-on-insulator\cite{rizzo2023massively}, thin-film lithium niobate\cite{boes2023lithium}, and silicon nitride\cite{riemensberger2022photonic}. By integrating passive photonics elements with III-V semiconductor optical amplifiers, on-chip laser systems have already been realized in wavelength ranges optimized for telecommunications\cite{xiang2021laser, xiang20233d, zhou2023prospects}, and are now moving towards the near infrared (NIR) and visible range \cite{tran2022extending, corato2023widely, zhang2023photonic}. However, integrated semiconductor laser and amplifier systems face several intrinsic limitations, most notably limited gain bandwidth and poor high-power handling capabilities due to two-photon\cite{ahmad2008energy} and free-carrier absorption in high-confinement waveguides, limiting on-chip pulse energies to several picojoules \cite{chang2022integrated}. An emerging alternative to on-chip III-V systems are integrated solid-state gain media based on rare-earth ions. Recently, by implanting ultra-low-loss silicon-nitride waveguides with erbium ions, high-performance solid-state waveguide amplifiers and lasers have been demonstrated\cite{liu2022photonic,liu2023fully} exceeding 100~pJ pulse energies; however, these efforts, along with those in other rare-earth systems, are limited to wavelengths longer than $1$~$\mu$m. Thus, the aforementioned applications that require the high-performance capabilities of Ti:Sapphire lasers have not been able to benefit from photonic circuit integration. An integrated solid-state laser and amplifier operating both in the visible and NIR wavelength range would have a sweeping impact in areas of science and technology where cost, size, and scalability considerations are critical.

Accordingly, efforts towards miniaturization of Ti:Sapphire lasers have a long history.  Limitations in scalability for current Ti:Sapphire laser systems have been primarily due to the short fluorescence lifetime of the Ti$^{3+}$ ion, which, when combined with the large disparity between pump (490 - 532 nm) and lasing (650 – 1100 nm) wavelengths, increases system complexity and necessitates very high pump intensities before achieving appreciable amplification and lasing. To address this, approaches such as pulsed laser deposition of waveguides \cite{grivas2005single}, laser written waveguides\cite{grivas2018generation, grivas2012tunable}, optical fibers\cite{yang2019widely, wang2016laser}, and machined whispering-gallery mode lasers\cite{azeem2022ultra} have been used to reduce mode volume and lower the lasing threshold. A recent demonstration of bonding Ti:Sapphire to a silicon nitride photonic chip to achieve lasing via evanescent coupling represents the first nanophotonic Ti:Sapphire laser\cite{wang2023photonic}, but this approach poses challenges for achieving substantial gain due to limited modal overlap. As a result, a practical on-chip Ti:Sapphire laser system has not yet been realized. To date, the highest single mode output power in an on-chip Ti:Sapphire laser is only 40~nW\cite{wang2023photonic}, and a tunable laser has yet to be demonstrated. Furthermore, no attempt to realize an integrated Ti:Sapphire amplifier has been reported. 

In this work, we address these challenges and demonstrate chip-integrated Ti:Sapphire lasers and waveguide amplifiers that deliver the power, stability, and tunability relevant for practical applications in research and technology. Our approach is based on an architecture that attains the fundamental limit of photonic mode confinement and overlap. We develop a low-loss monocrystalline sapphire-on-insulator photonics platform (Q-factors $>10^6$ at 805~nm), enabling a sub-milliwatt threshold Ti:Sapphire microdisk laser. 
We then demonstrate the first integrated Ti:Sapphire broadband waveguide amplifier, realizing a record 64 dB/cm solid-state on-chip gain. The waveguide amplifier is capable of pulsed amplification up to 2.3~nJ, the first pulsed amplification on-chip below $1$~\textmu m, which furthermore exceeds the power handling of any nanophotonic waveguide amplifier by over an order of magnitude\cite{liu2022photonic, shtyrkova2019integrated} while maintaining a transform-limited pulse shape.
Finally, we realize the first tunable integrated Ti:Sapphire laser, with the widest tuning range for any fully-integrated laser\cite{corato2023widely, guo2022hybrid, guo2023band}, approaching that of state-of-the-art free-space external cavity diode lasers \cite{mandelis2022review}. To illustrate the immediate practical relevance of the nanophotonic Ti:Sapphire-on-insulator (Ti:SaOI) laser, we use it to perform solid-state cavity quantum electrodynamics experiments that leverage its high output power, $<0.1$~pm wavelength stability, 24.7~THz coarse tunability, and $36.4$~GHz mode-hop-free tuning range.

\section{Titanium:sapphire-on-insulator photonics platform} 

The Ti:SaOI fabrication process is illustrated in Fig.~1(a). A Ti:Sapphire wafer is bonded to a sapphire carrier wafer via a SiO\textsubscript{2} interface that acts as the buried oxide layer. After bonding, the Ti:Sapphire layer is thinned via mechanical grinding and polishing, followed by reactive ion etching to a target thickness of $<1$~\textmu m. Using this approach, wafer-scale production of other high-quality monocrystalline photonics platforms, such as silicon-carbide-on-insulator\cite{lukin20204h}, has already been demonstrated. The fabrication method does not constrain the thickness of the Ti:Sapphire layer, which can be precisely controlled via dry etching. Figure 1(b) shows a photograph of a completed Ti:SaOI microchip ($8\times0.5$~mm) resting on a cube of bulk Ti:Sapphire crystal. The magnified view of the chip is presented in Fig.~1(c), showing an array of Ti:SaOI lasers (described in section IV).

To assess the intrinsic photonic qualities of the Ti:SaOI platform, we fabricate whispering gallery mode resonators via a photoresist pattern transfer process that results in an ultra-low-roughness surface. The photoresist mask is photolithographically defined on 350 nm thick Ti:SaOI, and a subsequent reflow and reactive ion etch produces microdisk resonators with surface roughness of 1.6~\AA~RMS (Fig.~1(d)). The sample is mounted on a temperature-controlled stage, and the devices are optically interfaced via a tapered fiber. A scanning laser is used to characterize the resonators in transmission across the lasing wavelength range. Due to the four-level structure of the Ti\textsuperscript{3+} gain medium, the lasing transition of Ti\textsuperscript{3+}  in the absence of pumping has no impact on absorption in the emission band, permitting direct characterization of passive losses. We observed intrinsic quality (Q) factors up to 1.3 million at a wavelength of 805 nm (Fig.~1(e)) in 50~\textmu m diameter resonators, corresponding to propagation loss of 0.5 dB/cm.  By thermally tuning the microdisk, it is brought in resonance with the 532~nm pump laser. The output lasing signal from the device is characterized via a spectrometer. We observe single-mode lasing with a threshold of 290~\textmu W (Fig.~1(f)), a reduction by a factor of $22$ from previous demonstrations \cite{wang2023photonic, azeem2022ultra}. This is a consequence of the tight modal confinement, maximal modal overlap with high-quality gain material, and low-loss fabrication.

Direct pattern transfer from photoresist to Ti:Sapphire limits the resolution and sidewall angle of structures, precluding the realization of many integrated photonics functionalities. To this end, we develop a pattern transfer process based on a secondary hard mask layer consisting of 200~nm of chromium (deposited via electron-beam evaporation). Electron-beam photoresist (FOx-16, Corning) is used to define the pattern in the Cr mask (see Methods). This process results in waveguides with a near-vertical (11~deg) sidewalls and sub-100 nm minimum feature size (Fig.~1(g, h)). We note that this process is also compatible with deep UV photolithography for high-throughput fabrication\cite{tran2022extending}. This high-confinement, nearly single-mode waveguide geometry achieves a near-unity broadband overlap of the pump and lasing modes (99.5\% at 650~nm, 96.5\% at 1100~nm), a crucial advantage of the Ti:SaOI laser platform that unlocks the ability to reach the ultimate limit of efficiency and low-power operation in Ti:Sapphire lasers. Of critical importance is that the thin film of Ti:Sapphire preserves the quality of the bulk gain medium, as quantified by its quantum efficiency.  A reduction of the gain medium efficiency through introduction of non-radiative decay rates would be measured as a reduced excited state lifetime. We characterize the Ti:SaOI waveguide fluorescence in the time domain by exciting with a weak modulated pump and detecting emission rate using a superconducting nanowire single-photon detector (SNSPD), and observe a purely single-exponential decay with a lifetime of 3.09~\textmu s, consistent with lifetimes reported in literature for bulk Ti:Sapphire \cite{moulton1986spectroscopic}. We conclude that the Ti:SaOI platform features excellent metrics for laser fabrication, both in passive optical losses and in the active gain medium.

%
%
\begin{figure*}[t]
\includegraphics[width=\linewidth]{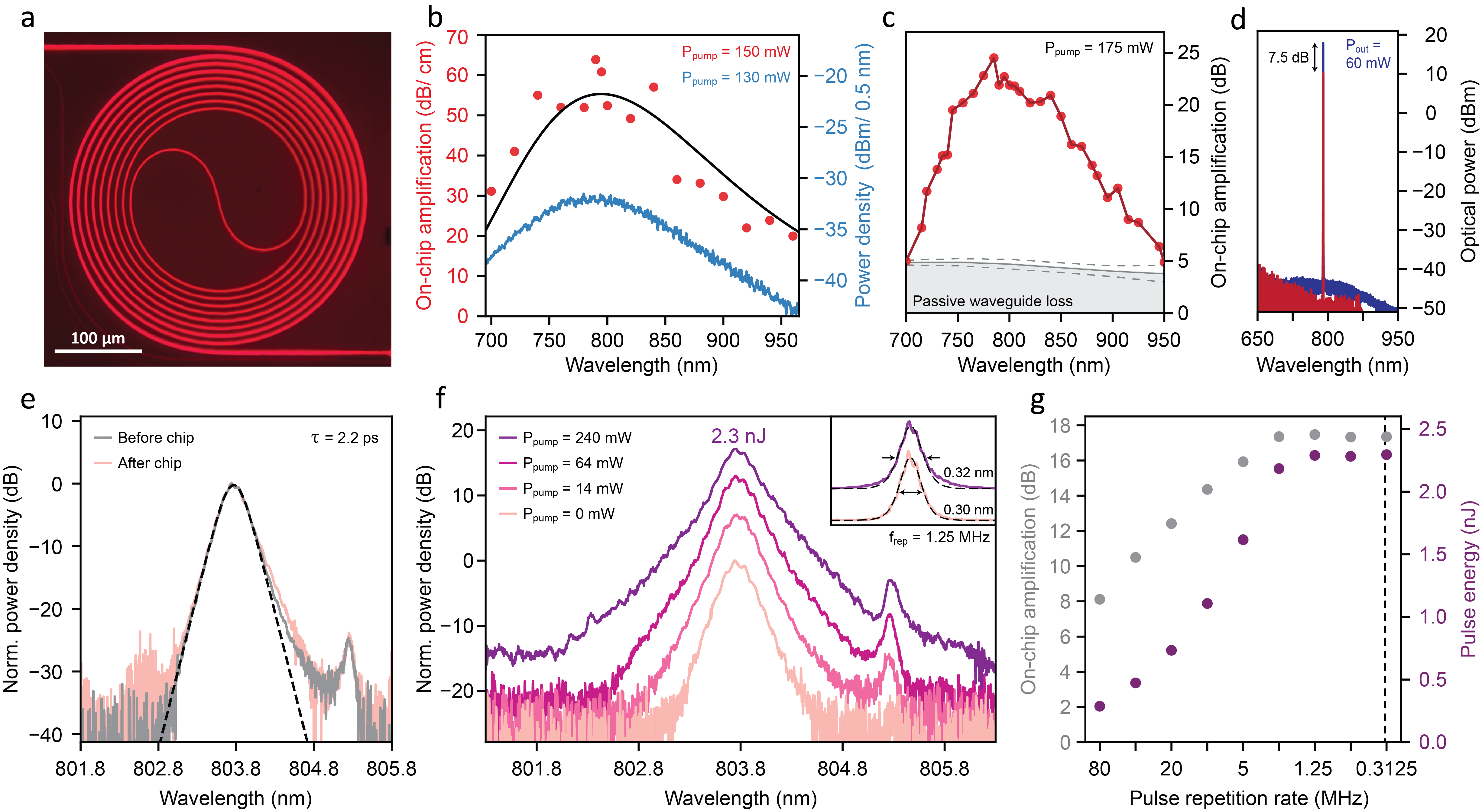}
\centering
\caption{\textbf{Integrated optical amplifier in Ti:SaOI.} 
(a) Optical image of an 8~mm waveguide amplifier pumped from both ends (532~nm pump filtered). 
(b) Amplified spontaneous emission (ASE) spectrum collected from a 3 mm waveguide amplifier when no input signal is present (blue). Measured small-signal gain (dB/cm) for a 0.5~mm waveguide (red). The fit to the calculated gain profile is shown in black\cite{burton2017temperature}. 
(c) Measured small-signal gain for an 8 mm amplifier (red), with pump power of 175 mW, and measured passive waveguide losses (grey). Standard error is shown via dashed lines.
(d) Spectrum of the amplified CW output signal at peak recorded output power of 60~mW.
(e) Normalized spectra of a picosecond pulse before and after propagation in the amplifier, showing no nonlinear distortion. 
(f) Spectrum of the amplified pulses (at repetition rate of 1.25~MHz) for varying magnitude of amplification. Inset shows unamplified and maximally-amplified output on a linear scale, illustrating that the the pulse remains transform-limited with minor distortion of the tails. 
(g) The dependence of amplification and output pulse energy on the repetition rate. Vertical dashed line indicates the lifetime of the gain medium ($1/\tau$).}
\label{fig_2}
\end{figure*}
\section{Ultra-wide band waveguide amplifiers in Ti:SaOI} 
Amplifiers are essential companions to lasers, allowing for reversal of losses in coherent signals and enabling the reach of power levels beyond the capability of a laser source. In absence of an input signal, broadband amplifiers also can generate incoherent light, powering technologies such as optical coherence tomography (OCT)\cite{bouma2022optical}. Solid-state amplifiers have excellent power-handling characteristics and their media feature low passive losses. Commercial solid-state amplifier solutions are mostly based on rare-earth dopants in crystals or glasses, which are limited to wavelengths longer than 1~\textmu m. The integration of such amplifiers on-chip via implantation of rare-earth ions into silicon nitride\cite{liu2022photonic}, lithium niobate\cite{jia2022integrated}, or alumina\cite{shtyrkova2019integrated} photonics is an area of active research, but is limited by the intrinsically-low gain of such dopants, posing a major technological challenge. To date, the realization of on-chip optical gain elements below 1~\textmu m has been limited to III-V semiconductors. Their heterogeneous integration with Si or SiN photonics has enabled breakthrough integrated laser technologies \cite{zhou2023prospects, tran2022extending, de2020heterogeneous, de2021iii, xiang2021laser}; however, limited power-handling and bandwidth constrain their applications and versatility. Using the Ti:SaOI platform, we demonstrate the first integrated solid-state amplifier operating below 1 \textmu m, which possesses unmatched power handling and bandwidth among both integrated and table-top state-of-the-art systems. 

Outside of nanophotonics, the use of Ti:Sapphire as a versatile amplifier is impossible, limited to costly, specialized applications such as ultra-high pulse-energy amplifiers. We demonstrate that strong pump confinement (effective mode area 0.62~\textmu m$^2$) and near-unity mode overlap in nanophotonics results in modest power requirements to reach saturation of the gain medium, enabling high gain even at low pump power. The Ti:SaOI waveguide amplifier (Fig. \ref{fig_2}(a)) is fully-integrated and ultra-compact ($0.20$~mm$^2$). In absence of an input signal, it is a source of bright amplified spontaneous emission with a bandwidth exceeding 100 THz (Fig. \ref{fig_2}(b)), the widest demonstrated to date. Such a broadband source, which can be readily coupled to an optical fiber, may find use in medical applications such as OCT\cite{drexler1999vivo, eggleston2019towards}. To measure the maximum gain achievable in Ti:Sapphire nanophotonics, we first examine a short (length of 485~\textmu m) waveguide, where effects of pump absorption, passive losses, and spatial mode mixing are negligible. A commercial continuous-wave (CW) Ti:Sapphire laser (SolsTiS, M Squared) is used as a wavelength-tunable signal. At a pump power of 150~mW, we observe peak gain of 64~dB/cm, and $>~30$~dB/cm across a 100~THz bandwidth (Fig.~2(b)). 

To achieve practical levels of amplification, we use a spiral waveguide with a length of 8~mm, where the pump is nearly fully absorbed. We note that, due to the four-level structure of Ti\textsuperscript{3+}, no signal re-absorption takes place in weakly-pumped sections of the waveguide, and material gain is always positive. The small-signal (3~$\mu$W input power) amplification of an 8 mm Ti:SaOI waveguide is shown in Fig. \ref{fig_2}(c), reaching peak on-chip gain of $>20$~dB. Here, the maximum gain was limited by parasitic lasing caused by reflections from the waveguide facets, similar to that reported for on-chip Erbium amplifiers in Ref.~\cite{liu2022photonic}. 
Higher gain can be achieved via a reflection-suppressing design or packaging. Figure~2(d) shows large-signal amplification, with an output CW power of 60~mW at 790~nm.

High-performance mode-locked lasers have long eluded photonic integration due to the challenge of amplifying ultra-short, high peak-power optical pulses. These lasers are pivotal for advancing diverse fields such as supercontinuum and optical frequency comb generation\cite{holzwarth2000optical}, two-photon microscopy\cite{helmchen2005deep}, and dual-comb metrology\cite{ideguchi2013coherent}. Here, we show that the combination of high gain-per-unit-length, absence of two-photon absorption, weak Kerr nonlinearity ($n_2 = 3\cdot10^{-20}$~m$^2$/W)\cite{major2004dispersion}, and excellent power-handling of Ti:SaOI enables unprecedented distortion-free pulsed amplification on-chip.  We use a commercial picosecond mode-locked Ti:Sapphire laser (Tsunami, Spectra Physics) as the input signal. To assess the degree of passive nonlinear distortion of a pulse propagating in the amplifier, we compare the pulse spectrum before and after passing through the waveguide, without amplification. Negligible spectral distortion is observed for an input pulse energy of 120 pJ (Fig.~2(e)). The shape of the output pulse for varying pump power is shown in Fig.~2(f), and shows minimal change in shape at peak gain of 17~dB, which corresponds to a pulse energy of 2.3~nJ and peak power of 1.0~kW. This is over an order of magnitude greater than the highest integrated pulsed amplification demonstrated to date, based on rare-earth waveguides \cite{liu2022photonic,shtyrkova2019integrated, byun2009integrated}, and is the only integrated high-power transform-limited amplifier for any wavelength. The Ti:SaOI amplifier constitutes the first realization of on-chip pulsed amplification below 1~\textmu m, and demonstrates the potential of the platform for the realization of a high-performance integrated mode-locked laser.

\begin{figure*}[t!]
\includegraphics[width=0.8\linewidth]{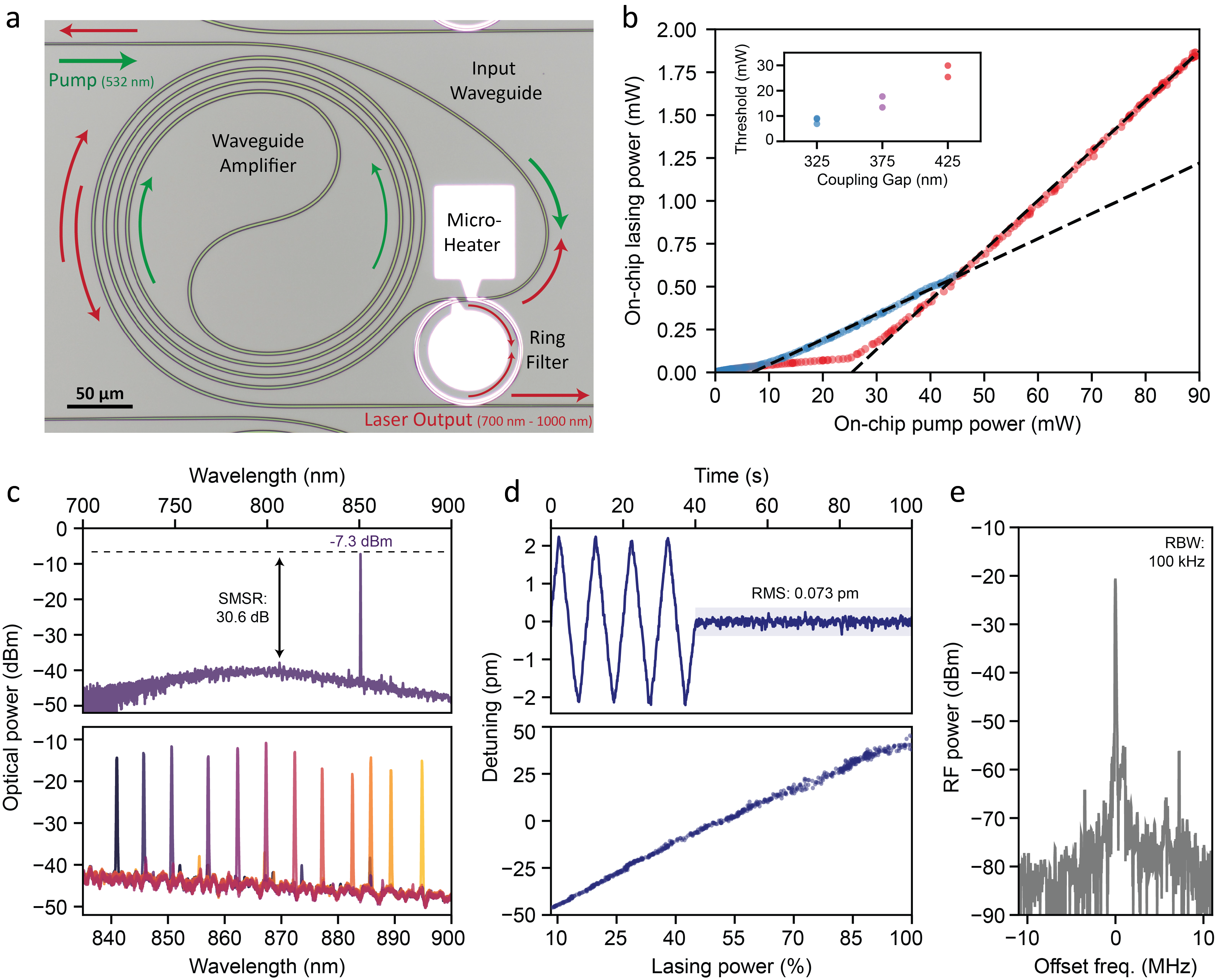}
\centering
\caption{\textbf{Widely-tunable, narrow-linewidth chip-integrated Titanium:Sapphire laser.} 
(a) Optical image of one laser.
(b) Output power of the Ti:SaOI tunable laser versus pump power (multimode operation) for the largest (red) and smallest (blue) coupling gap. Inset: dependence of threshold power on microring-waveguide coupling gap across the seven-device laser array.
(c) Top: Lasing spectrum and ASE background (here and onwards, operation is single-mode).
Bottom: Coarse wavelength tuning via the microheater.
(d) Top: Fine wavelength tuning and stabilization via feedback applied on the microheater.
Bottom: Mode-hop-free tuning by 91~pm (36.4 GHz) via modulation of pump power. 
(e) Measured laser linewidth of 140~kHz, via heterodyne technique.}
\label{fig_3}
\end{figure*}

\section{Tunable Ti:Sapphire laser on chip}

Owing to the four-level structure of Ti:Sapphire, the demonstrated high-gain waveguide amplifier straightforwardly enables the realization of a tunable, non-resonantly pumped single-mode laser when combined with a spectrally-selective recirculating element. In this section, we demonstrate a versatile Ti:SaOI laser based on this architecture. The laser schematic is presented in Fig.~3(a). The re-circulating element is a microring resonator integrated with a platinum microheater for spectral tuning, isolated from the gain section and thus avoiding temperature-induced gain degradation. For lasing to occur at a given wavelength, two conditions must be met: (i) the round-trip accumulated phase must be zero, and (ii) the microring must be on-resonance with the lasing wavelength (the transmission of the ring resonator must be sufficient for gain to overcome round-trip loss). This interplay of the delay line phase and microring resonance condition enables widely tunable, single-mode operation.  Furthermore, this geometry benefits from the disparate pump and signal wavelengths: Despite broadband coupling between the waveguide and microring in the lasing band, there is negligible coupling at the pump wavelength, and thus the architecture operates purely non-resonantly and maintains near-unity modal overlap (Fig.~1(i)). Thus insensitive to the spectral purity of the pump source, this laser can be pumped with any laser diode in the range 450-550~nm, as will be shown.

All lasers studied here feature a 3.2~mm waveguide section and an 80~\textmu m diameter microring filter. The distance between the waveguide and microring at the coupling point is varied across the laser array (see Methods). The smallest coupling gap achieves threshold as low as 6.9~mW; The largest gap yields the highest efficiency (2.9\%) and output power ($1.8$~mW, in multimode operation), as shown in Fig.~3(b). Figure~3(c) shows single mode operation across $>50$~nm wavelength range in a single device. Multimode lasing in the range from 790~nm to 930~nm has been observed among the devices studied in this work. To realize lasing across the full gain bandwidth of Ti\textsuperscript{3+} (650 - 1100~nm), an optimized filter-ring coupling geometry
will be necessary to compensate for the nonuniformity in the gain profile (Fig.~2(c)).

We investigate the spectral stability and mode-hop-free tunability of the laser via a high-precision wavelength meter (WS7, HighFinesse). By applying feedback to the microheater, we realize triangular tuning across 4~pm and wavelength stabilization to 0.073~pm RMS (Fig~3(d)). An even greater single-mode, mode-hop-free tuning range is possible via pump power modulation. Figure~3(d) shows mode-hop-free tuning of 91~pm (36.4 GHz), matching that of commercial Ti:Sapphire laser systems. Finally, we measure the instantaneous linewidth of 140 kHz via a heterodyne measurement against a commercial titanium sapphire laser (Fig.~3(e)).

For the results presented so far, we have used a table-top green laser (Verdi, Coherent) as the pump, due to its convenience and large available power. However, such pump lasers are bulky (10~kg excluding power supply) and expensive (\$30,000). Thus, chip-integration of the Ti:Sapphire laser without miniaturization of the pump does not achieve scalability. Due to the non-resonant, low-threshold Ti:SaOI laser design presented in this work, the spectral coherence of the pump is irrelevant, allowing us to use an off-the-shelf laser diode (OSRAM, retail price \$37 on www.mouser.com) as a miniature pump. The experimental details and lasing spectra are presented in Methods and as Extended Data Figure 1, respectively. This complete integrated system driven by a cheap incoherent light source is the decisive first step towards democratizing the Ti:Sapphire laser technology for real-world applications.
\section{Employing a Ti:SaOI laser for optical control of cavity-integrated artificial atoms}
\begin{figure*}[t]
\includegraphics[width=\linewidth]{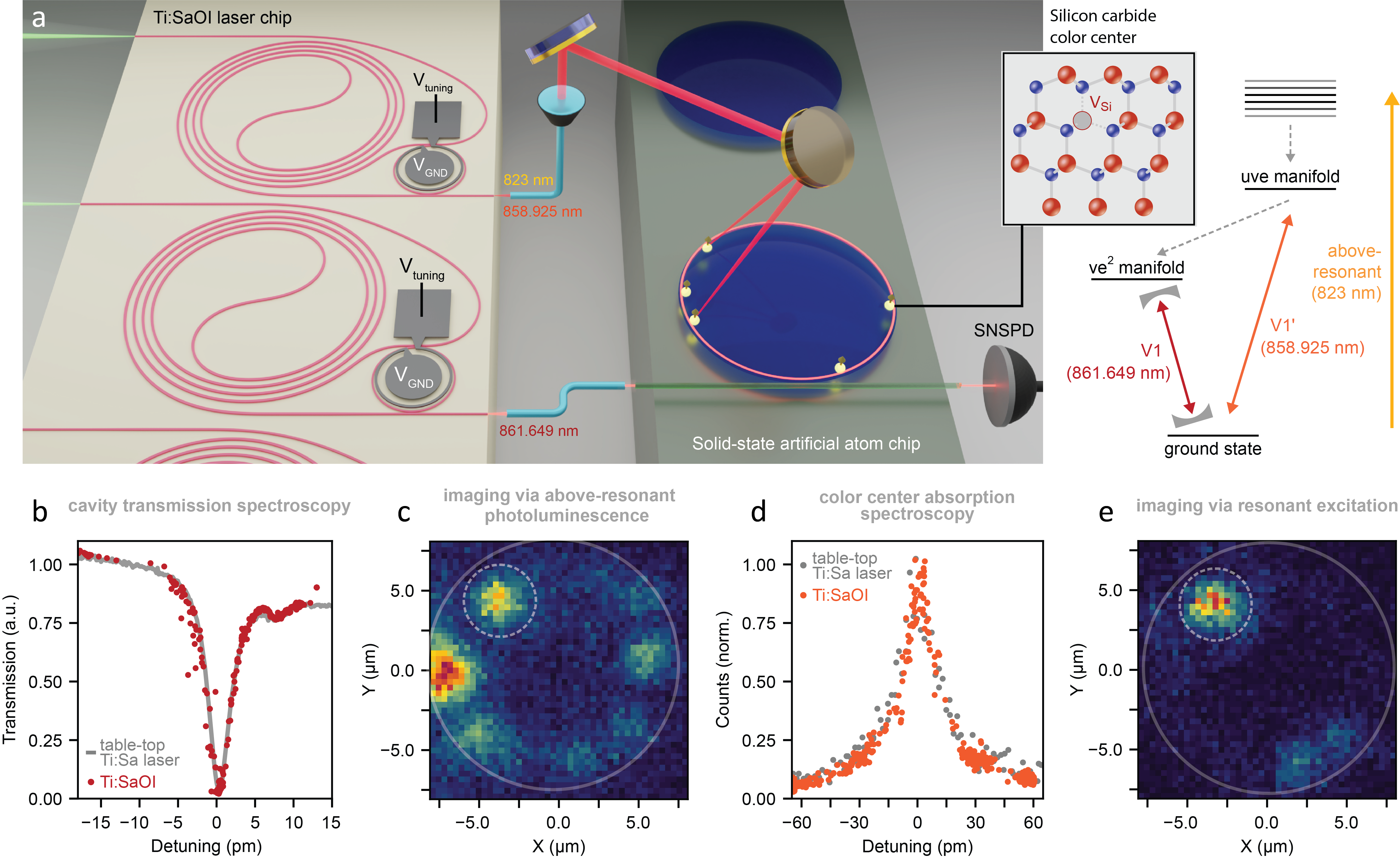}
\centering
\caption{\textbf{Quantum photonics with artificial atoms driven by integrated Ti:SaOI lasers.} (a) Illustration of the demonstration: A Ti:SaOI laser array is used to control a solid-state cavity QED system in silicon carbide (SiC). Lasers control the system via free-space excitation as well as via a waveguide interface. The level structure of the V\textsubscript{Si} color center is shown on the right. Panels (b-e) illustrate the flow of the cavity QED experiment: (b) The resonator frequency is tuned until an optical mode overlaps with the V1 transition of the color center, as confirmed via cavity transmission scan. Detuning is relative to 861.649~nm. For comparison, we show transmission recorded both with a table-top Ti:Sapphire laser and with our integrated Ti:SaOI laser. (c) With the cavity on resonance, an above-resonant raster scan is performed at 823~nm. A fluorescence spot (dashed circle) is identified as a color center candidate. (d) The free-space laser beam is focused on the spot, and resonant absorption spectroscopy is performed by tuning the laser across 120~pm to identify the V1' transition. Detuning is relative to 859.925~nm. (e) The raster scan is repeated with the resonant laser at the absorption line peak. The high contrast scan reveals that a single color center is coupled to the resonator, whereas the other features in (c) are parasitic fluorescence.}
\label{fig_4}
\end{figure*}

We proceed to demonstrate an application of the presented integrated Ti:Sapphire laser array in a practical laboratory setting: We perform a cavity quantum electrodynamics (QED) experiment with solid-state artificial atoms in silicon carbide (SiC), with the Ti:SaOI laser array as the sole light source.  Color centers in materials such as SiC and diamond \cite{awschalom2018quantum} can enable scalable, integrated, optically interfaced spin-qubits for advancing quantum sensing\cite{aslam2023quantum}, quantum networks\cite{bhaskar2020experimental}, and many-body physics\cite{davis2023probing}. Color centers are typically integrated into nanophotonic resonators in order to enhance their light-matter interaction, as well as to mediate atom-atom coupling to build entanglement\cite{evans2018photon, lukin2023two}. The operation of an atomic system requires multiple lasers with stable and precisely-tunable frequencies. Here, we replace all table-top lasers sources with a Ti:SaOI laser array.

The experiment is illustrated in Fig.~4(a). The output of two Ti:SaOI lasers on one chip is coupled into a single-mode fiber and routed to the cavity QED system. The cavity consists of a silicon carbide whispering gallery mode resonator \cite{lukin2023multiemitter} with a fiber interface\cite{catanzaro2023cryogenic}, cooled to 4.5~Kelvin in a closed-cycle Helium cryostat. The details of the experiment follow closely our previous work performed with table-top commercial lasers in Ref.~\cite{lukin2023two}. The system can be excited through the waveguide, or via free-space optical access from above, which enables diffraction-limited imaging of the color centers in the resonator via scanning beams. The artificial atoms used in this work are silicon vacancy (V\textsubscript{Si}) color centers. A simplified level structure of the V\textsubscript{Si} is shown in Fig.~4(a). The V\textsubscript{Si} features two optical transitions, around 861.6~nm and at 858.9 nm \cite{nagy2018quantum}, and is also amenable to above-resonant excitation with wavelengths below 830~nm. 

To achieve tunable operation at the precise wavelengths of V1 and V1' transitions, we operate in the few-mode (2 - 5 modes) lasing regime and filter the desired mode with an optical filter. In the future, more complex Vernier filter design will enable robust purely single-mode operation covering the entire tuning range without gaps. For above-resonant excitation where higher power is required but the wavelength does not need to be tuned precisely, we operate in single-mode regime. The experiment begins by tuning the resonator onto the V1 transition. This is done via cryogenic gas deposition while monitoring the waveguide transmission\cite{lukin2023two}. Figure~4(b) shows a critically-coupled optical mode of the cavity scanned with the Ti:SaOI laser (and a commercial laser for comparison). Once the cavity is on-resonance, the Ti:SaOI laser at 823~nm (output power 0.160~mW) is raster-scanned from free-space to detect the spatial distribution of emitters via waveguide-coupled photoluminescence (Fig~4(c)) with photon counting performed with an SNSPD. However, other parasitic optically-active defects are excited as well, and not all fluorescence spots correspond to desired artificial atoms. Ultimately, resonant excitation via V1' optical transition is the optimal means of excitation. To verify whether a fluorescent spot is a silicon vacancy color center, we park the Ti:SaOI laser at one spot and perform resonant absorption spectroscopy via continuous tuning of the laser over 160~pm across the V1' optical transition and detection of emission into the V1 transition. This measurement is shown in Fig.~4(d). Finally, we confirm that the other fluorescence in the above-resonant scan (Fig.~4(c)) does not correspond to silicon vacancy color centers by raster-scanning the laser beam parked at 858.925~nm, and observing no other dominant fluorescence spots (Fig.~4(e)). We have thus identified the spatial location and exact frequency of a color center after co-aligning it spectrally with the cavity to achieve Purcell enhancement, using an integrated Ti:Sapphire laser array. We note that, in this experiment, the Ti:SaOI laser was used in a plug-and-play configuration, replacing the table-top laser directly without need for modifications to the experimental setup. The demonstrated Ti:SaOI platform will permit simultaneous control of many emitters at disparate frequencies and play a key role in achieving scalability of both atomic and solid-state quantum systems.\cite{levonian2022optical}.

\section{Discussion}
In this work, we have demonstrated chip-integrated, broadband tunable, and scalable Ti:Sapphire lasers in a monocrystalline sapphire-on-insulator photonics platform. Single-mode operation across a 59~nm tuning range and single-mode output power of up to $-6.8$~dBm renders the demonstrated Ti:SaOI lasers immediately useful for practical applications. Already, without design optimization, lasers in this work match or exceed the mode-hop-free range of commercial solutions. The low threshold and non-resonantly pumped compact Ti:SaOI lasers  replace complex and expensive ($>$\$100,000) Ti:Sapphire laser systems, enabling a three orders-of-magnitude reduction in cost and size, as well as a two-orders-of-magnitude drop in power consumption of the full system, significantly lowering startup costs for many applications, such as atomic physics research. This will result in a democratization of high-performance Ti:Sapphire lasers, with a sweeping impact on research and technology. 

Beyond vastly increasing the accessibility of Ti:Sapphire lasers, Ti:SaOI also unlocks new modalities of this traditional laser crystal. Through nanophotonic confinement, highly-efficient generation of ultra-broad ASE shown in this work would serve as a versatile fiber-coupled source for high-resolution optical coherence tomography \cite{bouma2022optical}. The demonstration of a Ti:SaOI waveguide amplifier ($>$ 20~dB of on-chip gain, and peak gain of 64~dB/cm) opens doors to amplification of both continuous-wave and pulsed light within the ultra-wide bandwidth of 650 - 1100~nm. Owing to the excellent power-handling and weak nonlinearity of sapphire, we have shown transform-limited amplification of picosecond pulses up to 2.3~nJ pulse energies, constituting the first demonstration of integrated high-energy pulse amplification in the visible to near-IR and the first demonstration of pulse amplification exceeding 150~pJ in any nanophotonic waveguide amplifier \cite{liu2022photonic, shtyrkova2019integrated}. In comparison, this level of performance can never be attained with heterogeneously-integrated III-V semiconductor amplifiers due to their short upper-state lifetime\cite{girardin1998gain} and two-photon absorption\cite{ahmad2008energy}. This work addresses one of the major challenges in the development of on-chip mode-locked lasers and self-referenced optical frequency combs in a sought-after wavelength range previously out-of-reach\cite{newman2019architecture, spencer2018optical}. Beyond laser technologies, the monocrystalline sapphire-on-insulator platform shown in this work may find uses in demanding applications in the blue and ultraviolet range, owing to its ultra-wide bandgap of 10~eV and pristine crystal quality\cite{he2023ultra}.

The integration of Ti:Sapphire technology on-chip is a decisive step towards massively-scalable Ti:Sapphire systems, both monolithic and integrated with passive materials. Strides in heterogeneous laser integration have already enabled the combination of other novel on-chip laser technologies with large-scale foundry photonics processing\cite{tran2022extending, xiang20233d}. The approach presented in this work is compatible with co-integration in platforms such as silicon nitride and lithium niobate to realize sub-micron wavelength frequency-agile Ti:Sapphire lasers, integrated frequency-doubled lasers \cite{li2022integrated}, on-chip supercontinuum generation \cite{zhao2015visible}, and integrated optical parametric amplifiers\cite{ledezma2022intense}.

\noindent\textbf{Acknowledgments}\\
\noindent 
The authors would like to thank Carsten Langrock for his expertise in lapping and polishing, Kiyoul Yang for fruitful discussions and guidance on fiber tapering, Lavendra Mandyam for technical support in device fabrication, and Martin M. Fejer for access to laboratory equipment. The authors acknowledge funding support from the IET A.F. Harvey Prize, the Vannevar Bush Faculty Fellowship from the US Department of Defense, DARPA LUMOS, and the AFOSR under the award number FA9550-23-1-0248. J.Y. acknowledges support from the National Defense Science and Engineering Graduate (NDSEG) Fellowship. Part of this work was performed at the Stanford Nano Shared Facilities (SNSF)/Stanford Nanofabrication Facility (SNF), supported by the National Science Foundation under award ECCS-2026822.
\\

\noindent\textbf{Data Availability}\\
\noindent All data are available from the corresponding authors upon reasonable request.

\bibliography{main_ref}
\clearpage 

\section*{Methods}

\noindent\textbf{Device fabrication.} 
Thin-film Ti:SaOI is produced via an adaptation of the grinding-and-polishing approach for silicon-carbide-on-insulator platform described in Ref.~\cite{lukin20204h}. Multiple approaches and different crystal orientations are explored in this work: To produce the thin films used to demonstrate a microdisk laser, a c-plane (0001) Ti:Sapphire wafer die is bonded to an undoped sapphire wafer via an interfacial SiO\textsubscript{2} layer which serves as the buried oxide layer. The stack is then processed in a wafer grinder (DAG810, Disco Corp.), followed by chemical-mechanical polishing (POLI-400L from G\&P Tech.) and thinning via reactive-ion etching in BCl\textsubscript{3} plasma (PlasmaTherm Versaline ICP) to the target thickness of 350~nm. For subsequent amplifier and waveguide demonstration, we use an a-plane (1120) Ti:Sapphire wafer die, bonded to undoped sapphire via the same process. The bonded Ti:Sapphire dies are then ground and polished in a precision lapping system (PM5, Logitech), followed by the same reactive-ion etching process to reach a target thickness of 450~nm.

Microdisk resonators are fabricated via photolithography. Photoresist (S1822, Shipley) is spun on a Ti:SaOI film and patterned in a direct write lithography system (Microwriter ML3, Durham Magneto Optics). An optimized photoresist reflow process is then applied\cite{lukin2023multiemitter}, and the pattern is transferred to the Ti:Sapphire layer through a BCl\textsubscript{3} plasma reactive ion etch that minimizes surface roughness. Afterwards, the underlying SiO\textsubscript{2} is undercut in hydroflouric acid. 

Amplifiers and waveguide lasers are fabricated via electron-beam lithography. A chromium hardmask (200~nm) is deposited on the Ti:SaOI film by electron-beam evaporation. Chromium is chosen to allow for etching of sapphire with sufficient selectivity (4:1). Electron-beam photoresist (FOx-16, Corning) is spun afterwards and a 50 keV electon beam lithography system (Voyager, Raith) is used to define the pattern. Limitations of the lithography system necessitated confinement of devices to a $0.5\times0.5$~mm area to avoid stitching errors. The pattern is transferred into the Cr mask via a Cl\textsubscript{2} and O\textsubscript{2} plasma, and subsequently into sapphire via a BCl\textsubscript{3} etch. Subsequently, the Cr hard mask is stripped through wet etching via Chromium etchant and Piranha. Devices are then capped with an initial layer of flowable oxide (FOx-16, Dow Corning), followed by an additional SiO$_2$ layer through high-density plasma chemical vapor deposition (HDP-CVD, PlasmaTherm), and were annealed at 800$^{\circ}$C in air. Facets for edge-coupling of devices were created either through wafersaw dicing followed by focused-ion-beam milling or through laser stealth dicing.
\newline

\noindent\textbf{Device design.} Waveguide spirals for the amplifiers are designed following Ref. \cite{Chen:12} to minimize mode mixing during propagation. The spiral waveguides have a thickness of 450~nm and a width of 1.5~$\mu$m. Amplifiers with lengths of 3.4~mm, 8.1~mm, and 0.485~mm are studied in this work. The shortest amplifier is a straight waveguide with a width of 900~nm, extending across the chip.

The waveguide lasers use a spiral waveguide amplifier with a length of 3.2~mm, a thickness of 450~nm, and a width of 1.5~$\mu$m. A microring resonator with a diameter of 80~$\mu$m and width of 650~nm is used as the re-circulating element, and coupling gaps of 325~nm, 375~nm, and 425~nm were studied. The width of the ring waveguide was chosen such that a single transverse electric (TE) mode is supported at a given wavelength. A platinum microheater is patterned on top of the microring resonator for thermo-optic wavelength tuning. 
\newline

\noindent\textbf{Characterization of the waveguide amplifier.}
Light is coupled from free-space on- and off-chip via microscope objectives (M-plan NIR 100, Mitutoyo). The temperature of the sample substrate is maintained at 290~K throughout experiments. The 532~nm pump source is a frequency-doubled diode-pumped solid-state laser (Verdi V10, Coherent), and is delivered onto the chip either from one or both waveguide facets, or in one direction counter-propagating with respect to the signal. Unless otherwise stated, single-side pumping is used. For continuous-wave amplification experiments, a commercial Ti:Sapphire laser (SolsTiS, M Squared) is used, which has a tuning range of 700-1000~nm. Small-signal gain-per-unit-length is measured for an input power of 0.10~mW. For characterization of small-signal gain in the 0.485~mm long waveguide (Fig.~2(b)) requiring resolution of $<0.1$~dB, the pump and signal are chopped and the amplification is extracted from the time-dependent signal recorded on a silicon photodetector. For small-signal gain characterization in 8~mm amplifiers, an optical spectrum analyzer (Yokogawa AQ6370D) is used to obtain output signal with and without the pump laser. This measurement does not take into account the passive waveguide losses. Wavelength-dependent passive waveguide loss is measured from statistical analysis of waveguides of different lengths. Cumulative off-chip amplification is represented by the difference between measured on-chip amplification and passive loss at the given wavelength in Fig.~2(c). To record maximum amplified power output (Fig.~2(d)), the double-side pumped scheme is used. To infer the output power at the chip, the measured output power is corrected for objective losses (transmission of 0.85). For the pulsed-amplification experiment, a picosecond mode-locked Ti:Sapphire laser (Tsunami, Spectra-Physics) is used. The sech\textsuperscript{2} fit to the spectrum of the pulse reveals a pulse duration of 2.2~ps, in agreement with autocorrelator measurement. To measure the dependence of amplification and output pulse energy on the repetition rate, the intrinsic repetition rate of the laser (80~MHz) is reduced via an electro-optic modulator pulse picker (Conoptics) with a measured rejection ratio of $6.2\cdot10^3$. 
\newline

\noindent\textbf{Waveguide laser measurement}
Laser characterization experiments use the same setup described in the waveguide amplifier section, with the addition of electrical microprobes to deliver power to the microheater. For precise and accurate measurement of the lasing wavelength, a wavemeter is used (WS7, HighFinesse). Using real-time wavelength reading, a feedback loop is implemented to tune and stabilize the laser within a mode-hop-free tuning range of the microheater (Fig.~3(d)). The heterodyne linewidth measurement was performed with a commercial Ti:Sapphire laser (SolsTiS, M Squared). The photodetector signal was analyzed with an electrical spectrum analyzer (Agilent PXA N9030A) with a resolution bandwidth of 100 kHz.

The demonstration of Ti:SaOI laser in diode-pumped operation was performed as follows. A single-mode green laser diode with maximum output power of 110~mW was used (PLT5 520B from OSRAM, in a TO-56-3 package).
The laser diode was installed in a TE-Cooled Mount (LDM56, Thorlabs ), and powered by a laser diode driver (LDC210C, Thorlabs).
The laser diode output was collimated and coupled to the Ti:SaOI chip via free-space. Single-mode lasing at two wavelengths of operation are demonstrated by tuning the microheater (Extended Data Figure 1). The efficiency of pump coupling to chip was limited to $\sim$16\%, but can be improved substantially with mode overlap engineering and efficient direct-contact coupling, as demonstrated previously for silicon nitride photonics.\cite{corato2023widely} This approach can enable tunable Ti:SaOI chip-scale lasers in a butterfly package, as has been previously demonstrated in other platforms\cite{van2023long}.
\newline

\clearpage

\setcounter{figure}{0}
\renewcommand{\figurename}{Extended Data FIG.}
\renewcommand{\thefigure}{\arabic{figure}}

\begin{figure*}[t!]
\includegraphics[width=0.7\linewidth]{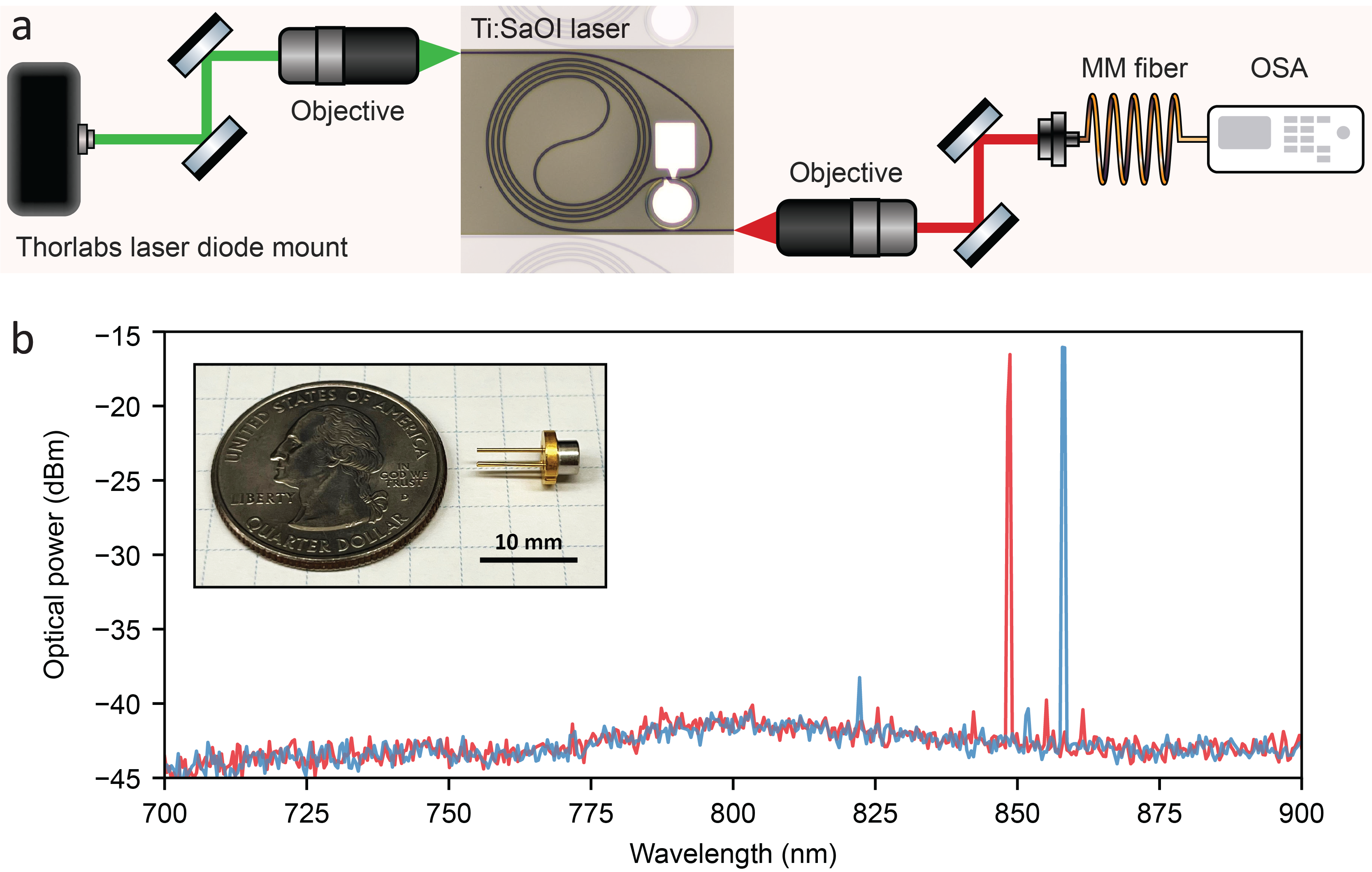}
\centering
\caption{\textbf{Diode-pumped on-chip Ti:Sapphire laser}
(a) Diagram of the measurement setup used in the diode-pumping experiments (MM: multi-mode, OSA: optical spectrum analyzer). 
(b) Measured optical spectrum of single-mode lasing at 848.7~nm and 858.3~nm, with a SMSR of 23.2 dB and 22.2 dB, respectively. (Inset) Image of the diode package used in these experiments.}
\label{fig_ext}
\end{figure*}

\pagebreak
\end{document}